\documentclass[aps,superscriptaddress,amsmath,showpacs,twocolumn,prl]{revtex4-2}
\usepackage[T1]{fontenc}
\usepackage[latin9]{inputenc}
\usepackage{amssymb}
\usepackage{graphicx}
\usepackage{subfigure}
\usepackage{epsfig}
\usepackage{txfonts}
\usepackage{amsfonts}
\usepackage{mathrsfs}
\usepackage{dcolumn}
\usepackage{esint}
\usepackage{mathrsfs}
\usepackage{enumerate}
\usepackage[colorlinks,
linkcolor=blue,
anchorcolor=black,
citecolor=blue,]{hyperref}%

\begin{document}

\preprint{APS/123-QED}

\title{Switchable coherent state stored quantum batteries with large ergotropies}

\author{Y. F. Li}
\altaffiliation[]{These authors contributed equally.}

\author{S. R. He}
\altaffiliation[]{These authors contributed equally.}

\author{P. H. Ouyang}
\author{X. N. Feng}%
\author{L. F. Wei}\email{lfwei@swjtu.edu.cn}
\affiliation{Information Quantum Technology Laboratory, School of Information Science and Technology, Southwest Jiaotong University, Chengdu 610031, China}

\date{\today}

\begin{abstract}
Quantum battery (QB) is a conceptually new energy storage and conversion device, which consists usually of a quantum charger and an energy store (called usually as the QB for simplicity). The demonstrated advantage of QB, over its classical counterpart, is that its charging efficiency can be significantly enhanced by using quantum entanglement resources. In this letter, we investigate alternatively how to realize the switchable charging and the lossless power detection of the charged QB. With the proposed QB configuration we show that, by adjusting the eigenfrequency of the qubit-based charger, the cavity-based QB can be switched on between the charging and power-off states and its stored energy can be non-destructively monitored by probing the transmitted spectrum of the external electromagnetic waves scattered by the qubit-based charger. As the qubit-based charger has never been excited, the proposed cavity-based QB could directly store the coherent state energy (rather than the single-photon one) and thus possess significantly large ergotropy. The physical realization of the proposed switchable QB configuration is demonstrated specifically with the experimental circuit quantum electrodynamical devices.
\end{abstract}

\maketitle

{\it Introduction.---}
Energy is one of the hottest topics in modern society, wherein various novel battery technologies including typically large-capacity supercapacitors~\cite{2022Yadlapalli}, solar photovoltaic cells~\cite{2021Jeong}, and wireless chargings~\cite{2016Eaton}, etc., have been paid much attention for the implementation of high-efficiency energy storage and conversions.
Interestingly, it has been demonstrated that quantum battery technology~\cite{2018Ferraro,2022Shi}, based on quantum physical principles, could be utilized to improve charging efficiency~\cite{2023Yang,2020Garciapintos,2018Niedenzu}.
Indeed, quantum coherence and entanglement could be regarded as important resources to push the development of information processing and energy technologies~\cite{2020Francica,2022Shi}.

Physically, an experimental quantum battery (QB) configuration, which can be simply represented as a charger-battery one, should consist of a quantum charger and energy store. Basically, they can be generated by various quantum systems, typically including natural/artificial atoms and high-quality electromagnetic cavities~\cite{2021Santos,2019Zhang,2019Santos}. Until now, a series of quantum charger-battery configurations, i.e., the atom-cavity~\cite{2022Shaghaghi,2023Salvia,2018Ferraro}, the cavity-atom~\cite{2023Almeida}, the cavity-cavity~\cite{2023Downing}, and the atom-atom ones~\cite{2022Dou}, etc., have been proposed to realize the desirable QBs. Importantly, it has been shown that the charging efficiency of the QBs could be significantly enhanced by using quantum entanglement resources. Particularly, unlike the atomic battery, the high-quality cavity-based battery can not only store energy for a long time~\cite{2023Salvia,2022Shaghaghi,2023RodriguezPRA,2021Blais, 2013Li} but also supports multiple electromagnetic modes for high capacity energy storage~\cite{2008Hofheinz}.

Basically, the quantum charging process of a QB is implemented by using quantum coherent evolution related to the charger-battery coupling, which can be either direct~\cite{2022Shaghaghi} or indirect (e.g., by eliminating their common couplings thermal reservoirs and thus could be non-reciprocal~\cite{2024Ahmadi, 2024Ahmadi}). However, these charger-battery couplings, up to our knowledge, are usually always on, yielding uninterrupted energy exchanges between the chargers and batteries. It would also limit the further improvement of the charging efficiency and the ergotropy of the charged battery~\cite{2021Blais}, in principle. A natural question is, does there exist a feasible approach to implement the switchable QB configuration for the switchable charging on demand?

In this letter, we propose a feasible QB configuration that utilizes a frequency-tunable qubit, as the quantum charger, to realize the switchable energy transfers from the external resource into the high-capacity QB for storage. Importantly, after the QB is fully charged, such an effective coupling between the cavity and the external driving field could be switched off, by simply adjusting the eigenfrequency of the qubit-based charger. In this case, the qubit-based charger could serve as a detector to non-destructively monitor the stored energy in the battery. 
Compared to the previous single-photon charging schemes~\cite{2023Downing}, the maximum extractable power (i.e., the ergotropy of QB) in the present switchable QB configuration can be significantly large (as the charger has never been excited~\cite{2023Rodriguez}), and thus the external coherent-state field energy can be directly transferred into the QB for storage.
Furthermore, beyond the usual resonant chargings~\cite{2021Santos,2019Zhang,2019Santos,2022Shaghaghi,2023Salvia,2018Ferraro,2023Almeida,2023Downing,2022Dou,2023Salvia}, the qubit-mediated configuration presented here can be used to implement the non-resonant, even the large detuning chargings. Therefore, more frequency resources could be utilized to implement the quantum chargings, although the frequency of the electromagnetic wave stored in the cavity-based battery is given physically.
Besides, as the present charging process can be conveniently switched off (whether the external field exists or not), the excessive heating of the QB (due to the always-on chargings) can be effectively avoided.
Particularly, the switchable QB configuration proposed here could be experimentally demonstrated with the well-developed solid-state quantum electrodynamical devices~\cite{2020Huang,2021Blais} and thus is scalable~\cite{2006Rabl,2008Goppl,2004Blais}.

\begin{figure}[htbp]
\setlength{\abovecaptionskip}{0.cm}
\setlength{\belowcaptionskip}{-0.cm}
\centering
\includegraphics[width=6.7cm]{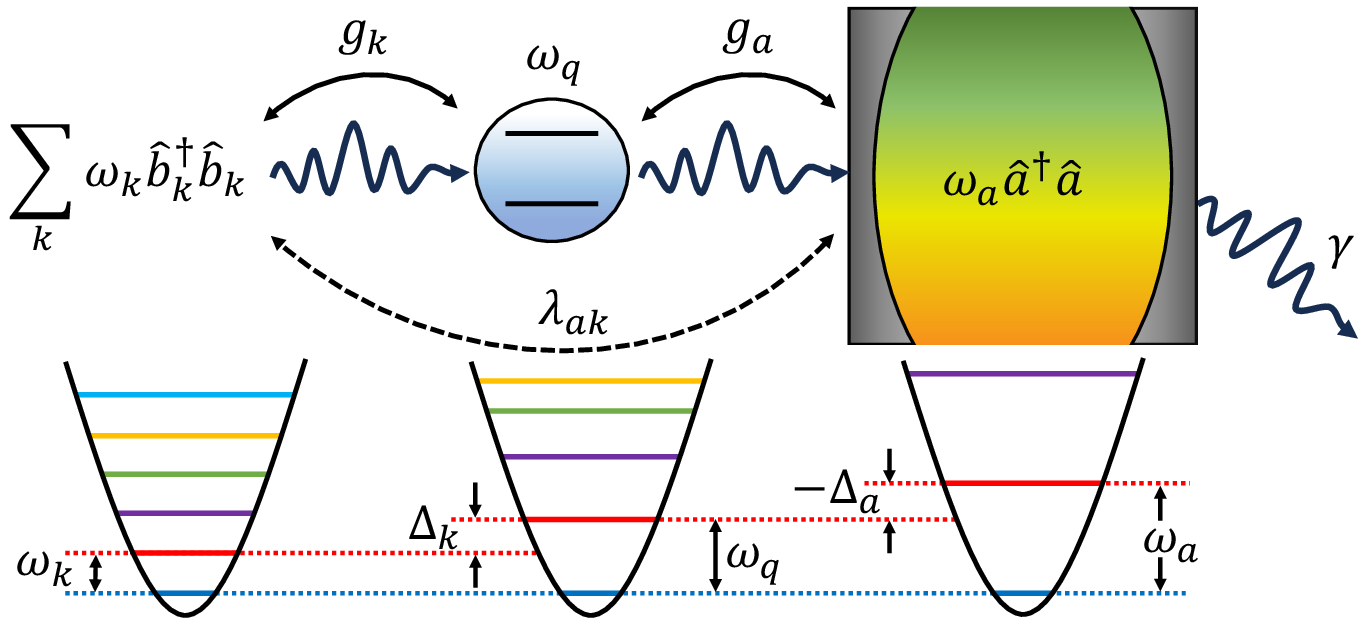}
\caption{(color online). A switchable QB configuration for transferring the $k$th mode external EMW (with the frequency $\omega_k$) into the cavity-based QB (with the frequency $\omega_a$) for storage. Here, the qubit with adjustable eigenfrequency $\omega_q$ acts as a quantum charger to implement the switchable charging and its detector to non-destructively monitor the charged power of the QB. $\Delta_{k,a}=\omega_q-\omega_{k,a}$, $\lambda_{ak}$ is the (indirect) effective coupling strength between the $k$-th mode EMW ($\hat{b}_k,\hat{b}^\dagger_k$) and the QB cavity ($\hat{a},\hat{a}^\dagger$) with the dissipation rate $\gamma$. $g_k$ and $g_{a}$ are the coupling strength between the qubit and external EMW and the cavity-based QB, respectively.}
\label{system}
\end{figure}

{\it Switchable QB configuration and its ergotropy.---}
Consider the QB configuration shown in Fig.~\ref{system}, wherein the high-quality transmission line resonator is served as the QB, the superconducting qubit generated by a SQUID-based superconducting circuit acts as a charger for transferring the electromagnetic wave (EMW) energy into the cavity QB for storage. The Hamiltonian of the system reads: $\hat H=\hat H_0+\hat H_I$, where $\hat H_0=\hbar\omega_q\hat \sigma_z/2+\hbar\omega_a \hat a^{\dagger} \hat a+\sum_k \hbar\omega_k \hat b^{\dagger}_k \hat b_k$ are the free Hamiltonians of the qubit-based charger (with the adjustable eigenfrequency $\omega_q$), the high-quality cavity (with the frequency $\omega_a$), and the external multi-mode electromagnetic field (with the frequency $\{\omega_k,\,k=1,2,3,...\}$) is used to drive the charger, respectively.
\begin{equation}
\hat H_I=\hbar g_a\left(\hat a^\dagger \hat \sigma^- + \hat a \hat \sigma^+\right)+\sum_k\hbar g_k \left(\hat b_k^\dagger \hat \sigma^- +\hat b_k \hat \sigma^+\right)   \label{eq1}
\end{equation}
describes the interactions between the charger and the cavity (with the strength $g_a$) and the $k$-th mode external field photons (with the strength $g_k$), respectively. Above, $\hat{a}$ ($\hat{a}^\dagger$) and $\hat{b}_k$ ($\hat{b}^\dagger_k$) are the annihilation (creation) operators of the cavity and external field, respectively. $\hat \sigma_z$ is the Pauli operator of the qubit-based charger.
In the dispersive regimes, i.e., for $g_a/|\Delta_a|$, $g_k/|\Delta_k| \ll 1$ with $\Delta_a=\omega_q-\omega_a$ and $\Delta_k=\omega_q-\omega_k$, the Hamiltonian of system can be effectively simplified as
\begin{equation}\label{HH}
	\hat H_{eff}=\hbar(\omega_a-\chi_a)\hat a^{\dagger} \hat a+ \hbar(\omega_b-\chi_{b}) \hat b^{\dagger} \hat b-\hbar\lambda_{ab}\left(\hat a\hat b^\dagger +\hat a^\dagger \hat b \right),
\end{equation}
see Supplemental Material~\cite{SM} A for details. Here, $\omega_b$ is the center frequency of externally charging field, $\chi_a=g_{a}^2/\Delta_a$, and $\chi_{b}= g_b^2/\Delta_b$, $\Delta_b$ and $g_b$ are the detuning and  coupling strength between the qubit and external field, respectively.
Obviously, the effective coupling  $\lambda_{ab}=g_ag_b(\Delta_a+\Delta_b)/\Delta_a\Delta_b$ between the driving external field and the cavity-based QB field is tunable, i.e., the charging process can be switched on/off for charging/stop on demand, by just adjusting the eigenfrequency $\omega_q$ of the qubit-based charger.

\begin{figure}[htbp]
\setlength{\abovecaptionskip}{0.cm}
\setlength{\belowcaptionskip}{-0.cm}
\centering
\subfigure{\includegraphics[width=4.4cm]{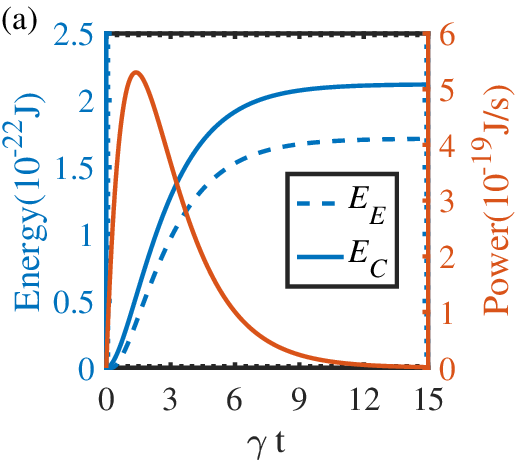}}
\subfigure{\includegraphics[width=4.05cm]{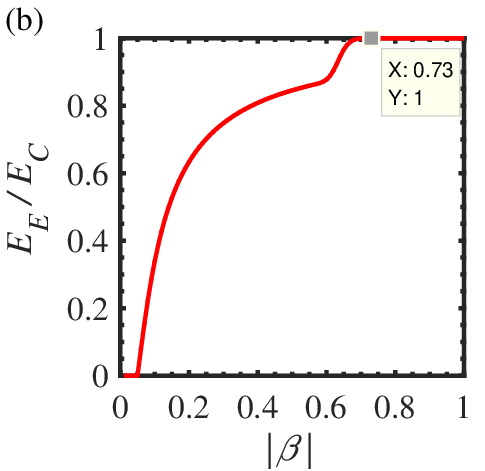}}
\caption{(color online). (a) The time-dependent average charging power (red line), the charged energy ($E_C$, blue solid line), and the ergotropy ($E_E$, blue dotted line) of the QB being at the charging state. (b) The ratio between the ergotropy $E_E$ and the charged energy $E_C$ versus the intensity $|\beta|$ of the coherent state EMW. The relevant parameters are set as: $\omega_a=2\pi\times5$ GHz, $\lambda_{ab}=0.1$ MHz, $|\beta|=0.4$ and $\gamma=0.01$ MHz.}
\label{power}
\end{figure}

First, let us discuss how to charge the cavity QB for energy storage. It is seen from Eq.~(\ref{HH}) that, once $\Delta_a+\Delta_b\neq 0$, the external EMW energy can be transferred directly into the fields in cavity-based QB, as the qubit-based charger has never been excited. This is different from the previous atom-cavity and cavity-cavity QB configurations, wherein the single-photon energy of the external field should be first charged into the charger and then transferred into the QBs for storage, which practically consumes the additional quantum coherent resources. To implement the high power charging, we demonstrate here that the QB configuration could be used to implement the coherent state charging, i.e., the externally driving EMWs are assumed to be at the usual singled-mode coherent state $|\beta\rangle$. Indeed, with the coherent-state driving external  field, the effective Hamiltonian~(\ref{HH}) consequently reduces
$
\hat H_{\beta}=-\hbar\lambda_{ab}(\hat a\beta^* +\hat a^\dagger \beta) =-\hbar\lambda_{ab}|\beta|(\hat a e^{-i\theta_b}+\hat a^\dagger e^{i\theta_b})$,
with $\beta=|\beta| e^{i\theta_b}$. Due to the practically existing dissipation of the cavity-based QB, such a charging process should be described by the master equation
\begin{equation}
\frac{d\hat \rho}{dt}=-\frac{i}{\hbar}\left[\hat H_{\beta},\hat \rho\right]+\frac{\gamma}{2} \left(2\hat a\hat \rho \hat a^\dagger-\hat a^\dagger \hat a \hat \rho-\hat \rho \hat a^\dagger \hat a\right).\label{drho}
\end{equation}
Solving this equation, we get $\langle \hat a^\dagger \hat a \rangle (t)=4\lambda^2_{ab} |\beta|^2(1-2e^{-\gamma t/2}+e^{-\gamma t})/\gamma^2$ and thus 
$
\langle \hat a^\dagger \hat a \rangle_{max}=4\lambda^2_{ab} |\beta|^2/\gamma^2$
for the steady state (i.e., $t\rightarrow\infty$). 
In Supplemental Material~\cite{SM} B, we further proved that the energy of EMWs is stored in the QB as the standing-wave coherent state $|\alpha\rangle$, with $|\alpha|=2\lambda_{ab}|\beta|/\gamma$. As a consequence, the average charging power for the duration $t>0$ can be easily calculated as
\begin{equation}
P_a^{(b)}(t)=\hbar \omega_a \frac{d\langle \hat a^\dagger \hat a\rangle }{dt}
=\frac{4\hbar \omega_a\lambda_{ab}^2 |\beta|^2}{\gamma}\left(e^{-\frac{\gamma}{2}t}-e^{-\gamma t}\right).
\end{equation}
Fig.~\ref{power}(a) shows how the charged energy versus the charging duration (the red line). One can see that, at the beginning (e.g., $\gamma t\leq 1.34$) the charging power increases significantly. Then, with the stored energy increases (e.g., $\gamma t>1.34$) the average charging power $P_a^{(b)}(t)$ gradually decreases, until it approaches zero for $\gamma t\geq 12$. When the stored energy (the blue line) approaches its maximum ($\sim 2.12\times 10^{-22}J$), i.e., the QB is almost fully charged, the charging process should be switched off. Fig.~\ref{power}(b) shows once the input power satisfies a certain condition, specifically for $|\beta|\geq 0.73$, the present coherent state charging schemes can provide the maximal ergotropy, i.e., all the charging energy can be extracted to do work.

We next investigate the ergotropy of the charged QB for the different charging durations. Following Ref.~\cite{2020Francica,2024Campaioli}, the maximum extractable work (i.e., ergotropy) of the QB prepared at the state $\rho(t)$ can be calculated as
$
\mathcal{E}(t)= Tr[\hat \rho(t)\hat H_B]-Tr[\hat \sigma_{\rho}\hat  H_B]$, where $\hat H_B=\hbar \omega_a\hat{a}^{\dagger}\hat{a}$ for the present cavity-based QB, $\hat \sigma_{\rho}(t)$ is the passive state of $\hat \rho(t)$. For the charging duration $t$ the state of the present cavity-based QB reads (see gain Appendix B): $\hat \rho(t)=|\mathcal{A}(t)\rangle\langle \mathcal{A}(t)| $, where $|\mathcal{A}(t)\rangle =\sum_{n=0}^{N} c_n(t)|n\rangle$, 
$
|\mathcal{A}(t)|^2=4\lambda^2_{ab} |\beta|^2 (1-2e^{-\gamma t/2}+e^{-\gamma t})/\gamma^2
$, and 
$c_n(t)=\mathrm{e}^{-|\mathcal{A}(t)|^2/2}[\mathcal{A}(t)]^n/\sqrt{n!}$.
Arranging the superposed Fock states (i.e., the eigenstates of $\rho(t)$) in an ascending order: $\left\lbrace |0\rangle,~|1\rangle,~|2\rangle,...,~ |N\rangle\right\rbrace $, and the eigenvalues $z_n(t)$ of the density $\rho(t)$ in descending order: $\left\lbrace z_0(t)=\max\left\lbrace c_n\right\rbrace ,~z_1(t),~z_2(t),...,~z_N(t)=\min\left\lbrace c_n(t)\right\rbrace\right\rbrace$ with $z_n(t)>z_{n+1}(t)$, we have $\hat \sigma_{\rho}(t)=|\zeta(t)\rangle\langle \zeta(t)|$, $|\zeta(t)\rangle=\sum_{n=0}^{N} z_n(t)|n\rangle$. Then, the ergotropy of the QB after the charging duration $t$ can be calculated as
\begin{equation}
\mathcal{E}(t)=\frac{4\hbar\omega_a\lambda^2_{ab} |\beta|^2}{\gamma^2}\left(1-2e^{-\frac{\gamma}{2}t}+e^{-\gamma t}\right)-\hbar\omega_a\sum_{n=0}^{N}\left|z_n(t)\right|^2,
\end{equation}
with $z_0(t)=0.2234$, $z_1(t)=0.2231$, $z_2(t)=0.2219$, and $z_3(t)=0.2212$ ..., for the typically parameters~\cite{2013Li,2019Koolstra}, e.g., $|\beta|=0.4$, $\gamma=0.01$ MHz, $\lambda_{ab}=0.1$ MHz, and $\gamma t=15$. Fig.~\ref{power}(a) shows how the ergotropy of the QB increases with the charging duration (the blue dotted line). It can approach a very large value, e.g., $51.73\hbar\omega_a$ (much larger than the single-photon energy) at the steady state. This is because the external EMW energy is directly transferred into the present QB and the qubit-based charger (which has never been excited) just serves as a data bus without any energy loss. This is different from the previous QB configuration proposed in Ref.~\cite{2023Downing}, wherein the additional dissipation exists in the charger generated by the charging cavity. Therefore, the present QB configuration can possess a significantly large ergotropy, as the coherent state energy (instead of the single-photon one) is transferred into the QB for storage. 

Secondly, let us discuss the issue of battery aging, i.e., its natural dissipation problem, after the QB is charged fully and the charging process is switched off. With Eq.~(\ref{HH}) one can see that if the eigenfrequency of the qubit-based charger is adjusted as: $\Delta_a=-\Delta_b$, we have $\lambda_{ab}=0$ and thus the QB decouples from the externally driving field. In this case, the charging process of the QB is stopped, although the EMW energy stored in the cavity-based QB must be lost. This is an aging process of the QB and is described simply by the following master equation
$
d\hat \rho/d\tau=\gamma (2\hat a\hat \rho \hat a^\dagger-\hat a^\dagger \hat a \hat \rho-\hat \rho \hat a^\dagger \hat a)/2   
$, for the aging duration $\tau$.
Beginning with the fully charged state, i.e., the QB is initially at the coherent state with the maximum average photon number $4\lambda^2_{ab} |\beta|^2/\gamma^2$, we simply have
$\langle \hat a^\dagger \hat a\rangle_d(\tau)=4\lambda^2_{ab} |\beta|^2e^{-\gamma \tau}/\gamma^2$. During such a natural aging, the state of the QB is changed as: $\hat{\rho}'(\tau)=|\alpha(\tau)\rangle\langle \alpha (\tau) |$ with $|\alpha(\tau)\rangle=\sum_nc_n'(\tau)|n\rangle$ and $|\alpha(\tau) |^2=|\alpha|^2e^{-\gamma \tau}$. By the same method described above, the ergotropy of the aging QB, after the decay duration $\tau$, can be calculated as
\begin{equation}
\begin{aligned}
	\mathcal{E}'(\tau)= &Tr\left[\hat{\rho}'(\tau)\hbar\omega_a \hat{a}^{\dagger}\hat{a}\right]-Tr\left[|\zeta'(\tau)\rangle\langle \zeta'(\tau)|\hbar\omega_a \hat{a}^{\dagger}\hat{a}\right]\\
	=&\hbar\omega_a\left(|\alpha|^2e^{-\gamma \tau}-\sum_{n=0}^{N}n\left|z_n'(\tau)\right|^2\right),
\end{aligned}
\end{equation}
with $
|\zeta'(\tau)\rangle=\sum_{n=0}^{N} z_n'(\tau)|n\rangle$ and $z'_0(\tau)=1$, $z'_1(\tau)=4.4\times10^{-3}$, and $z'_2(\tau)=1.38\times10^{-5}$..., specifically for $\gamma\tau=15$. Certainly, if the QB is required to be kept in the fully charged state for overcoming its natural aging, the charging process should not be switched off. This is the same as for the usual classical batteries.

\begin{figure}[htbp]
\setlength{\abovecaptionskip}{0.cm}
\setlength{\belowcaptionskip}{-0.cm}
\centering
\includegraphics[width=5.6cm]{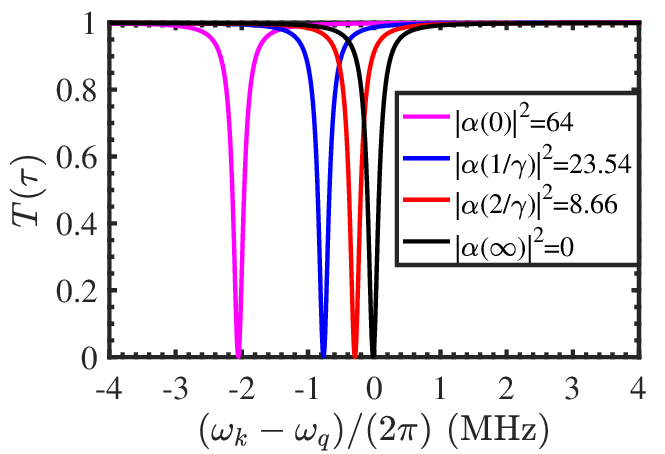}
\caption{(color online). The transmitted spectra of the external EMWs scattered by the dispersively coupled qubit-based charger, for non-destructively detection the photon number stored in the QB for different aging durations: the magenta, blue, red, and black lines correspond to the aging duration $\gamma\tau=0,1,2$, and $\infty$, respectively. Here, $|\alpha(\tau)|^2$ refers to the remaining average photon number at the aging time $\tau$,  $\omega_a=2\pi\times5$ GHz, $\omega_q=2\pi\times4.5$ GHz, $g_a = 10$ MHz, $g_l=3$ MHz. $|\beta|=0.4$ and $\gamma=0.01$ MHz.}
\label{power2}
\end{figure}

Additionally, for the practical application, the remaining power of the charged battery is often expected to be losslessly detected and displayed. This function can also be achieved with the present QB configuration. In fact, if only the qubit-charger is largely detuned from the cavity-biased QB, i.e., $g_a/\Delta_{a}\ll 1$, the charging process can also be switched off, as there is not any energy exchange between them. In this case, the qubit-based charger (without any dissipation ) serves just as the elastic scatter of the driving external field. Physically, such a scattering process can be described by the Hamiltonian~\cite{2005Shen,2020Gao}
\begin{equation}
	\begin{aligned}
		\hat H_2=&\hbar(\omega_a-i\gamma)\hat a^\dagger \hat a+\hbar\omega_q\hat \sigma^\dagger\hat \sigma^-+\hbar \int dx\tilde{b}_{R}^{\dagger}(x)(-iv_{g})\frac{\partial}{\partial x}\tilde{b}_{R}(x)\\&+\hbar\int dx\tilde{b}_{L}^{\dagger}(x)(iv_{g})\frac{\partial}{\partial x}\tilde{b}_{L}(x)+\hbar\frac{g_a^2}{\Delta_a}\hat a^\dagger \hat a\hat \sigma_z\\
		&+\sum_{j=R,L}\hbar\int dx g_l \delta(x-x_0)\big[\tilde{b}_j(x)\hat \sigma^++\hat \sigma^-\tilde{b}_j^\dagger(x)\big],
	\end{aligned}
\end{equation}
(see Supplemental Material~\cite{SM} C for detail).
Here, $\tilde{b}_{j}^\dagger(x)$ $(\tilde{b}_{j}(x))$ is the creation (annihilation) operator of the right $(j=R)/$left ($j=L$) traveling-wave photons scattered by the qubit-based charger and their coupling strength is $g_l$, $v_g$ is the group velocity of the driving external EMWs. Without loss of generality, the qubit-based charger is assumed to be at the location $x_0=0$, and thus the generic wave function of the present elastically scattering system can be written as 
$
|\psi(\tau)\rangle=|\alpha(\tau)\rangle\otimes\left\{\int dx[A_R(x)\tilde{b}_R^\dagger(x)+A_L(x)\tilde{b}_L^\dagger(x)]|\emptyset\rangle+A_e(x)\hat \sigma^+|\emptyset\rangle\right\},
$
where $|\alpha(\tau)\rangle$ is the decaying coherent state of the aging QB, $|\emptyset\rangle$ indicates that there are no photons in the driving electromagnetic field and the qubit is in the ground state, $A_{L/R}(x)$ and $A_e(x)$ denote the probability amplitude of a left/right traveling photons of the driving electromagnetic field and the excitation probability amplitude of the qubit, respectively. Solving the relevant Schr\"odinger equation for such a scattering process, we get the probabilistic amplitude~\cite{2022He}
\begin{equation}
\mathcal{T}(\tau)=\frac{i\left[(\omega_k-\omega_q)-g_a^2/(2\Delta_a)-g_a^2|\alpha(\tau)|^2/\Delta_a\right]}{i\left[(\omega_k-\omega_q)-g_a^2/(2\Delta_a)-g_a^2|\alpha(\tau)|^2/\Delta_a\right]-g_l^2/v_g}
\end{equation}
of the transmitted photons with $|\alpha(\tau)|^2=\langle \hat a^\dagger \hat a(\tau)\rangle$ being the average number of photons in the aging QB. 
Typically, Fig.~\ref{power2} shows how the observable transmitted spectrum $T(\tau)=|\mathcal{T}(\tau)|^2$ of the transmitted photons relates to the aging average photon numbers stored in the QB. With such a technique, the energy stored in the QB, i.e., the average photon numbers for various decayed states (such as the fully-charged state $|\alpha(0)\rangle$ the decay-depend state $|\alpha(\tau)\rangle$, and the vacuum state $|\alpha(\infty)\rangle$, etc.) can be non-destructively monitored in real-time, for its potential application. Obviously, such a function can not be possessed by the external field directly driving chargings~\cite{2023Downing}. 

\begin{figure}[htbp]
\setlength{\abovecaptionskip}{0.cm}
\setlength{\belowcaptionskip}{-0.cm}
\centering
\includegraphics[width=7.5cm]{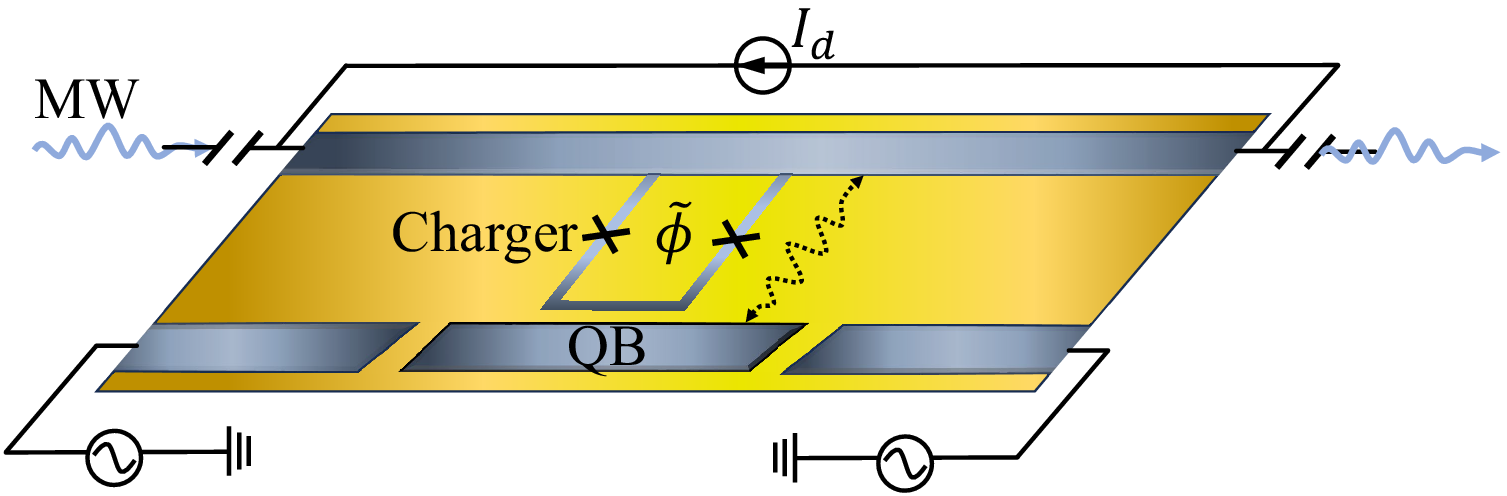}
\caption{(color online). A simplified circuit quantum electrodynamical circuit for experimentally realizing the proposed switchable QB. An artificial two-level atomic charger (generated by a double Josephson junction loop threaded by the magnetic flux $\tilde{\phi}$) is utilized to transfer the microwave (MW) energy of the transporting EMWs along the waveguide into the QB generated by a high quality factor transmission line resonator for storage. The eigenfrequency of the charger can be adjusted by varying the magnitude of the DC-biased magnetic flux $\phi_d$.}
\label{figPhysSys}
\end{figure}

{\it Possible physical realization.---} 
In general, the switchable QB configuration proposed here can be implemented by various solid- and optical systems, once their qubit-cavity interactions are switchable. Specifically, Fig.~\ref{figPhysSys} provides one of its physical realizations by using the well-developed circuit quantum electrodynamics system~(see, e.g.,~\cite{2008Wei}), wherein the high-quality transmission line resonator is served as the cavity-based QB for the EMW energy storage and the flux-modulated SQUID loop generates the qubit-based charger for transferring the external microwave energy into the QB.   

\begin{figure}[htbp]
\setlength{\abovecaptionskip}{0.cm}
\setlength{\belowcaptionskip}{-0.cm}
\centering
\subfigure{\includegraphics[width=4cm]{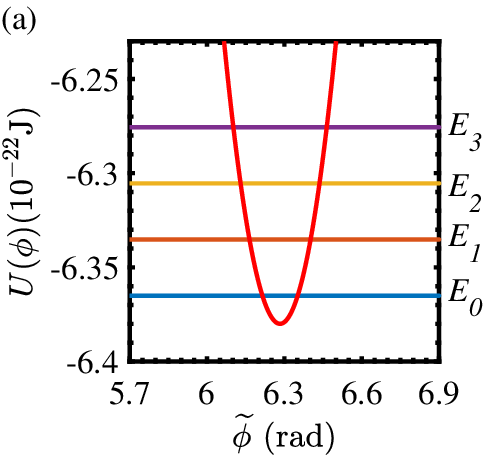}}
\subfigure{\includegraphics[width=4cm]{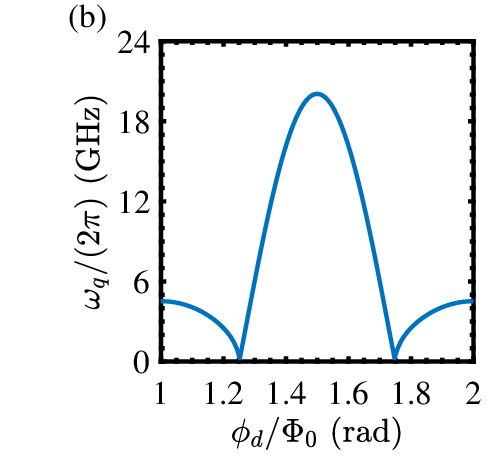}}
\caption{(color online). The energy level characteristics of the designed SQUID-based qubit. (a) Diagram of the qubit energy level structure when the magnetic flux is set to $\phi_d=1.977\Phi_0$ and $\tilde{\phi}_b=\tilde{\phi}_a=0.01\Phi_0$, $\omega_q=(E_1-E_0)/\hbar=2\pi\times4.5$ GHz, $E_0=-6.3814\times 10^{-22}$ J, $E_1=-6.3515\times10^{-22}$ J, $E_2=-6.3217\times10^{-22}$ J, $E_3=-6.2919\times10^{-22}$ J. (b) The qubit jump frequency is affected by the DC bias flux. Here, the parameters are set without loss of generality respectively~\cite{2010Harris,2024Ouyang}: $I_c=0.9794~\mu$A, $C=3.663$ pF.}
\label{figmodel}
\end{figure}

Physically, such an experimental circuit can be described by the following Hamiltonian  
\begin{equation}
\hat H_{\phi}=\frac{1}{4C}\left( \frac{2\pi}{\Phi_0}\hat p\right) ^2-2E_J\cos\left(\frac{2\pi}{\Phi_0}\hat \phi \right) \cos\hat \delta,\label{eq9}
\end{equation}
with $[\hat\delta,\hat p]=i\hbar$. Here, $\hat\delta$ is the gauge-invariant macroscopic phase operator of the SQUID loop with two identical Josephson junctions, $\Phi_0$ is the quantized flux, and
$\hat{\phi}=\phi_d+\hat{\phi}_b+\hat{\phi}_a$, $\tilde{\phi}=\phi_d+\tilde{\phi}_b+\tilde{\phi}_a$. $\phi_d$ being a large classical flux contributed by the biased DC-current, $\hat\phi_b=\tilde{\phi}_b(\hat{b}^\dagger+\hat{b})$ and $\hat\phi_{a}=i\tilde{\phi}_a(\hat{a}^\dagger-\hat{a})$ are contributed from the mutual inductance couplings of traveling-wave photons in the waveguide and standing-wave photons in the QB, respectively. 
$E_J$ is the Josephson energy of junction and $C$ is the total capacitance of the SQUID loop. With $\phi_d\gg\tilde{\phi}_b,\tilde{\phi}_a$, Eq.~(\ref{eq9}) can be expanded as $\hat{H}_\phi=H_{SQUID}+\hat{H}_I$, where $\hat H_{SQUID}=( 2\pi\hat p/\Phi_0)^2/(4C)-U_0(\phi_d) \cos\hat \delta$ generates an artificial atom with the potential $U_0\approx 2E_J\cos\left(2\pi\phi_d/\Phi_0 \right)$. The relevant bound-state levels of the superconducting artificial atom are shown in Fig.~\ref{figmodel}(a) for certain experimental parameters. While, $\hat{H}_I=-2E_J\sin(2\pi\phi_d/\Phi_0)\cos\hat{\delta}\,[\tilde{\phi}_b(\hat{b}^\dagger+\hat{b})+i\tilde{\phi}_a(\hat{a}^\dagger-\hat{a})]$
describes the couplings between the atom and the photons in waveguide and those in QB, respectively.  
Using the lowest two levels of the flux-biased SQUID loop to encode the superconducting qubit, one can see from Fig.~\ref{figmodel}(b) that its eigenfrequency is really adjusted in a sufficiently large regime. As a consequence, the Hamiltonian (\ref{HH}) required to generate the desirably switchable QB configuration, can be realized with the circuit, designed in Fig.~\ref{system}, with (see Supplemental Material~\cite{SM} D for details) 
\begin{equation}
\hat H_{I}'=i\hbar g_a\left(\hat a \hat \sigma^+-\hat a^{\dagger}\hat \sigma^-\right)+\hbar g_b\left(\hat b \hat \sigma^++\hat b^{\dagger}\hat \sigma^-\right).\label{eq8}
\end{equation}
Here, $g_a = 2\pi U_1\tilde{\phi}_a\mu_{01} /(\hbar\Phi_0)$, $U_1\approx 2E_J\sin\left(2\pi\phi_d/\Phi_0 \right) $, and $g_b = 2\pi U_1\tilde{\phi}_b\mu_{01} /(\hbar\Phi_0)$, with $\mu_{ij}=\langle i |\cos\hat \delta|j\rangle$ being the transition matrix element of qubit. This indicates typically that, with the circuit shown in Fig.~\ref{figPhysSys}, the switchable QB proposed above is really feasible on-chip.
Indeed, to implement the charging process, the experimental parameters can be set to satisfy the condition: %$\omega_a/(2\pi)=5$ GHz, $\omega_b/(2\pi)=4$ GHz, $\omega_q/(2\pi)=4.52$ GHz, yielding $g_b=38.9$ MHz, $g_a=9.77$ MHz,
$g_a/|\Delta_a|\approx3.2\times10^{-3}\ll 1$, $g_b/|\Delta_b|\approx0.01\ll 1$, and thus the effective coupling $\lambda_{ab}\sim1.52$ MHz for $\phi_d=1.992\Phi_0$. %can be obtained for effectively coupling the external field to the QB. 
For stopping the charging the QB, one can just adjust the dc biased flux as $\phi_d=1.977\Phi_0$ and get %$\omega_q=4.5$ GHz and thus
$\lambda_{ab}=0$. This indicates that the proposed QB device designed in Fig.~\ref{figPhysSys} is really switchable. 

{\it Conclusions and discussions.---}
Although most of the QB configurations proposed previously had demonstrated certain quantum advantages over the classical configurations, the always-on charger-QB couplings limit their conveniences for potential applications. To overcome such a difficulty, in the present work, we designed alternatively a switchable QB configuration, wherein a eigenfrequency-controllable qubit serves as the quantum charger for the tunable transferring the external coherent state energy into the high-quality cavity-based QB for storage. Compared with the other cavity-based QB, we demonstrated that the ergotropy of the present QB can be significantly large, as the EMW stored in the QB is still in its coherent state. Also, we showed that the remaining energy stored in the QB can be monitored non-destructively in real time by probing the transmitted spectrum of the external traveling photons scattered by the qubit-biased charger, which is dispersively coupled to the cavity-based QB.  

For the feasibility of the proposed QB configuration, we designed an experimental superconducting circuit based on the well-developed circuit quantum electrodynamical technique. As the generated superconducting qubit works in the microwave band, we argued that such a QB device could be utilized to implement the desirably fast wireless chargings, although certain techniques such as the working bandwidth, charging efficiency, and also a natural aging problem, etc., are still required to be well solved. Anyway, given energy crisis is a global problem, developing various quantum technologies, including the desirable QB ones, by using various quantum resources to improve energy storage, transfers, and also utilization is particularly desirable, at least theoretically. 

{\it Acknowledgment.---}
This work was partially supported in part by the National Key Research and Development Program of China under Grant NO. 2021YFA0718803, and the National Natural Science Foundation of China Grants NO. 11974290.

\end{document}